# The Second Law and Informatics

Oded Kafri

Varicom Communications, Tel Aviv 68165 Israel.

## Abstract

A unification of thermodynamics and information theory is proposed. It is argued that similarly to the randomness due to collisions in thermal systems, the quenched randomness that exists in data files in informatics systems contributes to entropy. Therefore, it is possible to define equilibrium and to calculate temperature for informatics systems. The obtained temperature yields correctly the Shannon information balance in informatics systems and is consistent with the Clausius inequality and the Carnot cycle.

# Introduction

Heat is the energy transferred from one body to another. The second law of thermodynamics gives us universal tools to determine the direction of the heat flow. A process is likely to happen if at its end the entropy increases. Similarly, energy distribution of particles will evolve to a distribution that maximizes the Boltzmann -$H$ function [1] namely, an equilibrium state, where entropy and temperature are well defined.

Information technology (IT) is governed by energy flow. Processes like data transmission, registration and manipulation are all energy consuming. It is accepted that energy flow in computers and IT are subject to the same physical laws as in heat engines or chemical reactions. Nevertheless, no consistent thermodynamic theory for IT was proposed. Hereafter, a thermodynamic theory of communication is considered. The discussion starts by drawing a thermodynamic analogy between a spontaneous heat flow from a hot body to a cold one and energy flow from a broadcasting antenna to receiving antennas. This analogy may look quiet natural. When a file is transmitted from a transmitter to the receivers, the transmitted file's energy, thermodynamically speaking, is heat. The Boltzmann entropy and the Shannon information have the same expression [2], so we can think about information increase in broadcasting with analogy to entropy increase in heat flow. However, to complete the analogy, it is necessary to calculate a temperature to the broadcasting antenna and the receiving antenna.

To establish this analogy one has to calculate, for informatics systems, the thermodynamic quantities appearing in the second law, namely, entropy, heat, and temperature, and to define equilibrium. These informatics- thermodynamics quantities should comply with the Clausius inequality [3,4] and to reproduce the Carnot efficiency.

The classical thermodynamics of heat transfer from a hot bath to a cold bath and the basic definitions of entropy, heat, temperature, equilibrium, and the Clausius inequality are a major part of the discussion of information thermodynamics. Therefore, in section **I** a brief review of these concepts is provided.

A file with a given bits distribution resembles a frozen two-level gas. The thermodynamics of a two-level gas in which the location of the exited atoms is constantly varying in time is well known. Deriving the thermodynamics of a file calls for comparison between the thermodynamics of a two-level gas and that of a binary file. In section **II**, a calculation of the entropy, heat, temperature, and the definition of equilibrium for the transfer of a two-level gas from a hot bath to a cold bath according to statistical mechanics are provided and the compliance with the Clausius inequality is shown.

Following this review, in section **III**, an analysis of the transfer of a binary file (a frozen two-level gas) from a broadcasting antenna (a hot bath) to receiving antennas (a cold bath) is provided. A temperature is calculated to the antenna that, together with the transmitted file information (entropy) and its energy (heat), is shown to be in accordance both with classic thermodynamics (i.e. the Clausius inequality) and information theory. The difference between randomness of a two-level gas, and randomness of a binary file and its effect on the thermodynamic quantities is discussed. It is concluded that the Shannon information is entropy.

Based on the results of section **III**, in section **IV** the second law of thermodynamics is defined for informatics. It is argued that reading/writing a file is equivalent to an isothermal compression/expansion of an ideal gas and amplifying/attenuating a file is equivalent to an adiabatic compression/expansion of an ideal gas. An ideal amplifier cycle comprises of two adiabatic and two isotherms is shown to have the Carnot efficiency.

Finally in section **V**, this theory is used to calculate a thermodynamic bound on the computing power of a physical device. This bound is found to be the Landauer's principle.

**I - Classical thermodynamics of heat flow**.

In this section a short review of the quantities that will be used later for Informatics systems is provided [3]. The second law of thermodynamics is a direct outcome of maximum amount of work $\Delta W$ that can be extracted from an amount of heat $\Delta Q$ transferred from a hot bath at temperature $T_H$ to a cold bath at temperature $T_C$ [3]. This amount of work can be calculated from the Carnot efficiency,

$$\eta \equiv \Delta W/\Delta Q \leq 1 - T_C/T_H. \tag{1}$$

Namely, the maximum efficiency $\eta$ of a Carnot machine depends only on the temperatures $T_C$ and $T_H$. To obtain the maximum efficiency the machine has to work slowly and reach equilibrium at any time in a reversible way. Clausius [4] defined the entropy, $S$, in equilibrium, such that it reproduces the Carnot efficiency, namely,

$$\Delta S \geq \Delta Q/T \tag{2}$$

If one dumps an amount of heat $\Delta Q$, to a thermal bath at temperature $T$, in a reversible way, the change in the entropy of the bath is $\Delta S = \Delta Q/T$, and the system is in equilibrium. If one dumps the heat irreversibly the system is not in equilibrium and $\Delta Q/T$ is smaller than $\Delta S$ as a result of efficiency lower then $\eta$. The entropy change $\Delta S$ is equal to $\Delta Q/T$ only in a reversible dumping in equilibrium. Therefore, if we assume that any system has a tendency to reach equilibrium, any system tends to increase $\Delta Q/T$. Taking a system out of equilibrium requires work, since the system will eventually reach equilibrium (namely, the energy of the work will be thermalized), therefore the entropy of a closed system tends to increase and cannot decrease. Temperature and entropy are defined in equilibrium and the temperature can be calculated as,

$$T = (\Delta Q/\Delta S)_{equilibrium} \qquad (3)$$

Note that away from equilibrium entropy and temperature are not well defined [3].

Consider a simple example of the entropy increase in heat flow from a hot thermal bath to a cold one (see Fig 1).

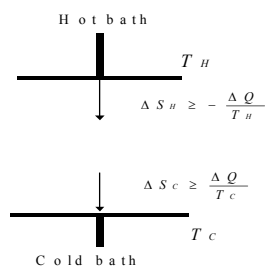

Figure 1: *The entropy increase in spontaneous energy flow from a hot thermal bath to a cold thermal bath.*

When we remove an amount of energy $\Delta Q$ from the hot bath, the entropy reduction at the hot bath is $\Delta Q/T_H$. When we dump this energy to the cold bath, the entropy increases by $\Delta Q/T_C$. The total entropy increase is $\Delta S = \Delta Q/T_C - \Delta Q/T_H$. One can see that if the process is not in equilibrium $\Delta S > \Delta Q/T_C - \Delta Q/T_H$. In general

$$\Delta S \geq \Delta Q/T_C - \Delta Q/T_H \qquad (4)$$

Hereafter, it is shown that inequality (4) is true both in statistical physics (sections **II**) and in information theory (section **III**).

**II - Statistical Physics of a two-level gas.**

A binary file resembles a two-level gas. However, in a two-level gas particles exchange energy and in a binary file the energy distribution of the bits is fixed. Hereafter, a thermodynamic analysis of a two-level gas transmitted from a hot bath to a cold bath is reviewed. In section **III** the thermodynamic quantities that will be calculated for file transmission will be compared to those of a two-level gas and the origin of the differences is discussed.

Boltzmann has shown that the entropy of a system can be expressed as

$S = -k \sum_{i=1}^{\Omega} p_i \ln p_i$ where $i$ index the possible microscopic configuration of the system, $p_i$ is the probability to be in the $i^{th}$ configuration, $\Omega$ is the total number configurations and $k$ is the Boltzmann constant. If all configurations are equally probable $p_i = 1/\Omega$, then $S = k\ln\Omega$, [3]. This expression will be used to calculate the thermodynamic quantities appearing in the Clausius inequality for a system that resembles an informatics system. Consider a thermal bath at temperature $T_H$, which is in contact with a sequence of $L$ states. $n$ of the $L$ states have energy $\varepsilon$ and will be called "one".

*L -n* of the states have no energy and will be called "zero". We analyze the thermodynamics of transferring this two-level gas from a hot bath at temperature $T_H$ to a colder bath at temperature $T_C$ with analogy to the heat flow analysis of section **I**. To calculate the entropy we need to count the number of configurations of the two-level sequence, namely, the possible combinations of *n*, "one" particles in *L* states. As can be seen in Fig 2 this number is the $n^{th}$ binomial coefficient. Namely, there are, $\Omega = L!/[\,n!\,(L-n)!]$ possible combinations.

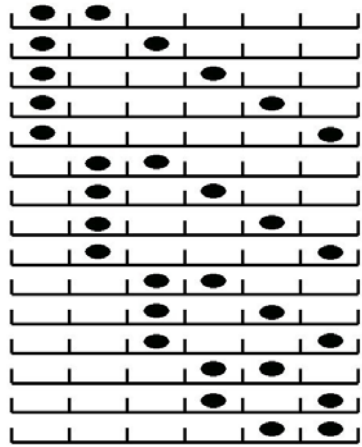

Fig 2: *A two-level gas with L=6 and n=2. In equilibrium all possible combinations have an equal probability. If some of the combinations have higher probability than others, the system is not in thermal equilibrium.*

The entropy of the system is $k \ln \Omega$, and the internal energy of the system is $U = n\varepsilon$. Since the gas is entirely removed from the thermal bath $\Delta Q = U$ and the temperature is given by Eq. (3). Using Stirlings formula we derive $\partial Q / \partial S$ to obtain *T* [5]. The well-known result is,

$$\frac{n}{L-n} = e^{\frac{-\varepsilon}{kT}} \quad \text{or} \quad T = \frac{\varepsilon}{k \ln \frac{L-n}{n}} \qquad (5)$$

Eq.(5) is the Maxwell Boltzmann distribution for a two-level gas [6]. For a given $\varepsilon$, one parameter $T$ represents all the knowledge on the two-level gas in equilibrium. This is a well-defined system with a well-defined entropy, temperature and energy. The equilibrium was invoked by giving an equal probability distribution to all the possible combinations $\Omega$ of the $n$ particles in $L$ states. If a system is not in equilibrium, there are certain combinations that are preferred and therefore the gas has a biased distribution. An unbiased distribution is the probability distribution, which describes the information we have about a system in the most honest way that allows us to make the best prediction about the property of a system. Jaynes has shown [7,8] that unbiased distribution yields the Shannon information. In a biased distribution the actual combination span is smaller, and $\Omega$ of the gas is smaller. Boltzmann called the quantity $k\sum_{i=1}^{\Omega} p_i \ln p_i \geq -S$ calculated for a biased distribution the $H$ function [1].

Hereafter, we calculate the entropy balance when a two-level gas is removed from a hot bath and is dumped into a cold bath for reversible and irreversible operation. It is shown that the process complies with the Clausius inequality

When the two-level gas is removed from the hot bath, the entropy is reduced by $\Delta S_H = n_H \varepsilon / T_H = k n_H \ln[(L-n_H)/n_H] \geq \Delta Q / T_H$. When we dump it to the cold bath, we generate an entropy $\Delta S_C = n_H \varepsilon / T_C = k n_H \ln[(L-n_C)/n_C] \geq \Delta Q / T_C$.

The total change in the entropy is,

$$\Delta S_C - \Delta S_H = \frac{k\Delta Q}{\varepsilon} \ln \frac{n_H(L-n_C)}{n_H(L-n_H)} \geq \frac{\Delta Q}{T_C} - \frac{\Delta Q}{T_H} \qquad (6)$$

If $T_C$ is lower than $T_H$, then $n_C > n_H$ and we see that eq. (6) is positive with accordance with eq. (4), namely, the Clausius inequality (see Fig 3).

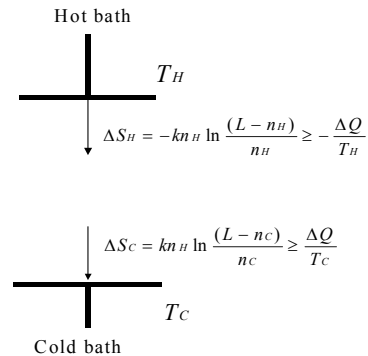

Fig 3. *The entropy increase, due to transmission of a two-level gas, from a hot bath to a cold bath is with accordance with the Clausius inequality.*

**III – Thermodynamics of information.**

In the Shannon model a binary file is transferred from a transmitter to a receiver. A binary file can be viewed as a frozen two-level gas. A binary file is not in thermal equilibrium as only one possible combination of the bits is transmitted.

Shannon's first theorem deals with the maximum amount of information that can be coded in a given binary file of length *L* in a noiseless channel and in a noisy channel (Shannon second theorem). This amount of information is called the Shannon information and it was shown to have the same expression as the Boltzmann entropy, namely $I = -\sum_{i=1}^{L} p_i \log p_i$ [2]. Many papers were written on the connection between the Shannon information and the Boltzmann entropy [8,9,10,11]. However in this paper a connection between Shannon information and the second law (i.e. the Clausius inequality) is discussed. The amount *n* of "one" bits, in a file of length *L*, has no unique relation to the amount of the Shannon information in the file. This is in contradistinction to a two-level gas in which the energy the temperature and the entropy are functions of *n* (see Fig. 3). For example, in a group of several files having the same *n* some may have a very small amount of Shannon information, i.e. when all the "one" bits are in the beginning of the file, and the rest of the file has zero bits or any other ordered combination (see Fig. 4) that can be regrouped effectively. Some other files may have a relatively high amount of Shannon information, if the distribution of the bits in the file is random as will be discussed later.

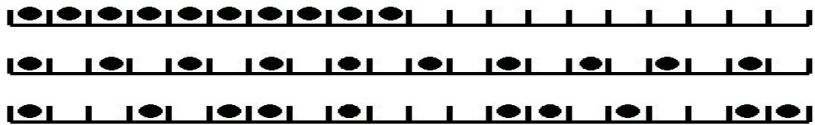

Fig 4: *Three possible binary files having the same energy. The higher two files have higher order and therefore contain little Shannon information. The lower file is random and is shown to be in equilibrium.*

The amount of the Shannon information in a file is a function of the randomness of the bits in it. The reason that Shannon obtained the same expression as Boltzmann is that, in a two-level gas in equilibrium we have no way to predict what combination of "one" particles will be at a certain time (see Fig.2), and in a random file we have no way to predict what bit will be at a certain time. The unpredictable sequence of bits is the useful information. When Alice is reading a binary file of length *L* she always obtains $L\ln 2$ useful nats (1bit = ln2 nat) even if the bits are ordered, because she lacks an *a-priori* knowledge of what bit will come next. However, Shannon first theorem is about Bob's ability to send a shorter file of length $L' \leq L$ from his knowledge of the sequence of the bits that he intends to send. The shortest file that can be recovered by Alice to reproduce the original file *L* is the amount of the Shannon information in the file. It is believed that the shortest file $L'$ is a random file (namely there is a conjecture that a random file cannot be compressed).

Let's examine an antenna broadcasting a file composed of energy and includes information, which is received by several antennas. Consider the radiating antenna as a hot bath emitting energy and entropy. Similarly consider the receiving antennas as a cold bath that absorbs energy and entropy. It is argued that if we assume that information is entropy, the information balance obeys the Clausius inequality. To calculate the thermodynamic functions of an informatics system one need to calculate a temperature for the antenna. We assume that the antenna's temperature is identical to that of the file that it emits or absorbs. This is with analogy to the two-level gas transmission that was discussed in the previous section.

The calculations of the thermodynamic properties of a file and a two-level gas are different. In two-level gas $\Omega$ is the number of combinations of $n$ particles in $L$ states, in a file $\Omega = 2^L$. In a two-level gas there is a well-defined ratio between the entropy and the energy that enable to calculate $T$, for any $n$, because randomness means an equal-probability for any combination of $n$ "one" particles in $L$ states. For a file, randomness means an equal-probability for any bit; therefore, a random distribution means $n \approx L/2$. If $n$ is not equal to $L/2$ there is no unique connection between the energy and the Shannon information and therefore a temperature cannot be calculated. For example, in Fig.4 all the three files have the same amount of energy. The upper two have very little information as Bob can compress them significantly. The lower file is a random file and contains more information. For a random file the ratio between the energy and the information is unique as $n = L/2$ and $\Delta I = L\ln2$, where $I$ is the Shannon information. So by assigning energy $\varepsilon$ to the "one" bit we obtain $\Delta Q = L\varepsilon/2$ and $\Delta S = k\Delta I = kL\ln 2$. Using Eq.(3) we obtain for the temperature,

$$T = \Delta Q/\Delta S = \varepsilon/k2\ln 2. \qquad (7)$$

In thermal systems, equilibrium is a state of randomness induced by collisions. Therefore, in analogy, it is assumed, that a random file is in equilibrium and has a well-defined temperature. This is in accordance with Clausius's result that temperature is defined only in equilibrium. The average energy per bit is $\varepsilon_n = \varepsilon/2\ln 2$, and therefore, eq.(7) yields that for informatics $\varepsilon_n = kT$.

The derivation of the temperature in eq.(7) is based on two major assumptions. The first one is that $\Delta S = k\Delta I$, namely, that the Shannon information is entropy as a consequence of the randomness of the bits in a file. The second assumption is that a random file is a state of equilibrium similarly to thermal systems in which randomness is a state of equilibrium. The obtained temperature $\varepsilon_n = kT$ is common in physics (i.e. harmonic oscillator). Nevertheless, it is necessary to show that these two assumptions encapsulated in the temperature of eq.(7) comply with the Clausius inequality and the Carnot efficiency.

The broadcasting of a file to several antennas is equivalent to heat flow from a hot bath to a cold bath. Specifically a broadcasting antenna, which broadcasts a file having a high-energy bit, is a hot bath. A receiving antenna, which absorbs a lower energy bit file, is a cold bath. The entropy multiplies according to the number of the receivers. To calculate the entropy-information balance we consider an antenna broadcasting a binary file to $N$ antennas. A possible realization of such system is a point-radiating antenna surrounded by a sphere, whose area is divided to $N$ equal receivers. The hot antenna emits the broadcasted file at a temperature $T_H$. A receiver antenna receives the broadcasted file with a lower temperature $T_C = T_H/N$. Since $\Delta Q/T = k\Delta I$, we obtain from eq.(4),

$$\Delta S \geq \Delta Q/T_C - \Delta Q/T_H = Nk\Delta I - k\Delta I. \qquad (8)$$

Eq. (8) shows that the file temperature, obtained in eq. (7), yields correctly the increase in information in the broadcasting of a binary file to $N$ receivers, which is $N\Delta I - \Delta I$. In fact the temperature is canceled out to give us simply the information balance from the thermodynamic quantities. In a "peer-to-peer" transmission, as in the Shannon model, no information increase is involved; therefore no thermodynamic considerations are necessary.

Now it is necessary to check if eq.(8) behaves according to the Clausius inequality out of equilibrium. Namely, if broadcasting of a none-random file yields the inequality sign. When there are correlations between the bits, the amount of the Shannon information in the file is smaller. As a result, the same energy carries less Shannon information i.e. $\Delta I$ is smaller than $\Delta S/k$. This shows that the second law of thermodynamics holds for informatics systems. Using eq.(8) we can rewrite the Clausius inequality for informatics system as, $\Delta S \geq k\Delta I$.

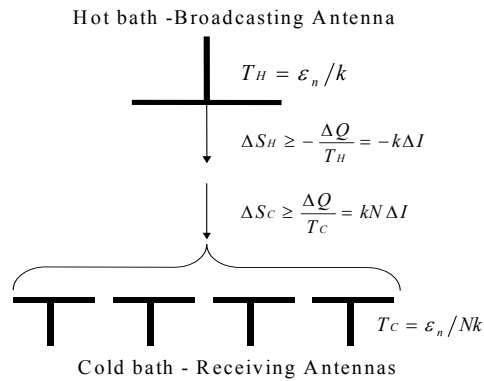

Fig. 5 *The analogy between heat flow from a hot bath at temperature $T_H$ to a cold bath at temperature $T_C$ and an antenna broadcasting a file of bit energy ε to N antennas, each receiving the file with bit energy ε/N. In the thermal case the entropy increase is $\Delta S \geq \Delta Q/T_C - \Delta Q/T_H$. However, the same equation $\Delta S \geq \Delta Q/T_C - \Delta Q/T_H$ reproduces well the information balance when we use the temperature definition from Clausius inequality $\Delta Q/\Delta S$ for a random binary file. The antennas deployment is drawn to emphasize the physics only.*

.

This implies that Information, like entropy, tends to increase. In a general case both informatics and thermal processes occurs simultaneously. In these cases a transformation of thermal entropy to informatics entropy and vice versa may occur. If we assume that entropy is an extensive quantity the Clausius inequality can be written as,

$$\Delta S \geq \Delta Q/T + k\Delta I \qquad (9)$$

The first term on the RHS is the thermal entropy and represents the ensemble randomness due collisions. The second term is the informatics entropy and represents a quenched randomness of the nats in a sequence (a file). The amount of the Shannon information in a partially random file, with some correlation between bits, is equivalent to the Boltzmann –$H$ function, namely the "entropy" calculated out of equilibrium. Shannon, in his famous paper [2], pointed out this analogy.

**IV The second law for informatics-the Carnot cycle**

In the previous section the energetic of a file broadcasted from one antenna to several antennas (a generalization of the Shannon theory) was studied. An analogy was drawn between information broadcasting from one antenna to several antennas to heat flow from a hot bath to a cold bath. What was shown is that:

1. The Shannon information content *I* of a file is equivalent to the Boltzmann -*H* function.
2. The transmitted file energy is equivalent to heat.

3. A random file is a state of equilibrium.
4. The temperature of the antenna is proportional to the average nat energy broadcasted from it or received by it.

These definitions comply with the Clausius inequality. We complete the analogy by demonstrating an informatics cycle, analogous to a Carnot machine.

The second law of thermodynamic is more renown in its verbal form, namely:

***It is impossible to construct a cyclic machine whose net outcome is transferring heat from a cold bath to a hotter bath.*** Namely, work has to be invested, from outside of a system, to transfer heat from a cold bath to a hotter bath. This definition of the second law is a direct outcome of the Clausius inequality. If one transfers an amount of heat $\Delta Q$ from a low temperature bath to a high temperature bath, $\Delta S$ is negative with a violation of the second law. Machines with a negative entropy balance are called *perpetuum mobile* of the second kind.

The Clausius inequality was deduced from the efficiency of the Carnot machine. The Carnot machine comprises of a cylinder equipped with a piston filled with an ideal gas. The Carnot machine transfers energy from a hot bath to a cold bath and produces work. The piston is first in contact with a hot bath at a temperature $T_H$. At the first stage the piston expands slowly at a constant temperature (isothermal expansion). In this stage the piston removes energy and entropy from the hot bath into the gas. In the second stage the piston is isolated from the hot bath and expands until the gas is cooled to the temperature of the cold bath. During this expansion the piston produces work against an external pressure. Since the cylinder is isolated, no heat is exchanged with the gas, so that its entropy remains constant (this process is called an adiabatic expansion). The third step is an isothermal compression; the gas in the cylinder dumps heat and entropy to the cold bath. The cycle is completed by an additional adiabatic compression of the gas to the temperature of the hot bath by applying work. The total work and energy balance yields the Carnot efficiency.

It is now shown that it is possible to construct an informatics Carnot cycle consists of two isotherms and two adiabatic. Reading a file is an analog of an isothermal expansion. When a file is received at constant bit energy, the energy of the receiver increases but its temperature remains fixed, exactly as in an isothermal compression. When a file is amplified, its temperature is increased but its information content remains fixed. This is an adiabatic compression.

An analog of a Carnot cycle is found in a transmitter of a file over an optical fiber. The file is amplified periodically at a given distances due to the signal attenuation caused by energy loss in the fiber. Carnot was interested in extracting mechanical work from a temperature gradient to convey physical goods over the friction of the railroad. In an optical fiber transmission, one wants to invest electrical work to covey information over the intrinsic loses of light on its path. The Carnot cycle for a file transmission over a fiber comprises of four steps identical to those of the original mechanical Carnot heat engine, see Fig.6.

1. A file is sent at high bit energy (high temperature) into the fiber. This is a writing process that is equivalent, as discussed above, to an isothermal expansion.

2. The file is transmitted through the fiber and during the transmission the file is attenuated. Its bit energy is reduced and the file is cooled. The information remains fixed and therefore this process is an adiabatic expansion. (In this example the work is lost, outside the informatics system).

3. An amplifier reads the file at a low temperature. During this process the energy of the amplifier is increased but its temperature remains constant. This is an isothermal compression.
4. The file energy is amplified at fixed information content. This is an adiabatic compression. The file is ready for a new cycle.

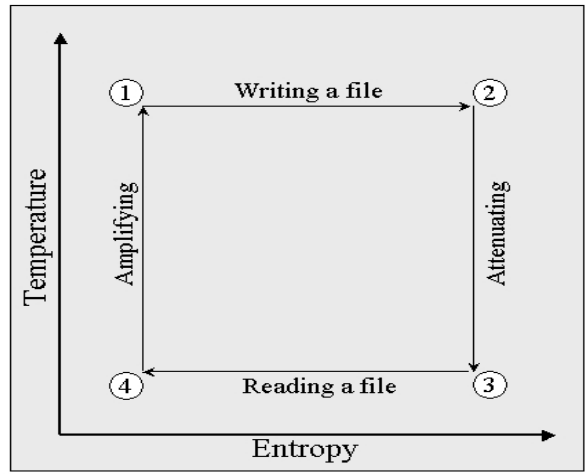

Fig. 6. *The Carnot cycle of a file transmission along an optical fiber. Amplifiers are necessary to overcome the energy loss in the fiber. Each cycle of amplification is shown to be a Carnot cycle of two isothermals and two adiabatics.*

With analogy to the original Clausius formulation it is possible to define the second law of thermodynamic for informatics:

***It is impossible to contract a cyclic amplifier whose net outcome is transferring heat from a low bit energy file to high bit energy file.*** Namely any amplifier requires outside work (i.e. a power source). If an amplifier receives a low temperature $T_C$ file and increases its temperature to $T_H$, at the end there is a negative entropy balance, $\Delta S = \Delta Q/T_H - \Delta Q/T_C < 0$, because higher the energy of a bit means less bits per $\Delta Q$. Therefore it requires adding to $\Delta Q$ an extra energy in order to avoid negative entropy. To conserve the entropy one needs that

$\Delta Q_H / T_H = \Delta Q_C / T_C$. Namely, the information of the hot file is equal to that of the cold file. Designating $\Delta Q_C = \Delta Q$ and $\Delta Q_H = \Delta Q + \Delta W$, where $\Delta W$ is the added work requires to avoid a negative entropy, we obtain that $\Delta W = \Delta Q (1 - T_C / T_H)$. Namely, the Carnot efficiency, of eq.(1).

This formulation is applicable as well to optics. Every picture is comprised of combination of spatial modes [11] i.e. pixels. These spatial modes are independent light sources. If one detects an image, for a given time period, it is possible to assign energy to the pixels of the image and thus to calculate a temperature in addition to the Shannon information content. Therefore we can generalize the 2$^{nd}$ law for optics;

***It is impossible to construct a passive imaging optical device that will produce an image with energy flux higher than that of the original image.***

The above phrases are by no means surprising or novel. However, it is shown that energy flow in computers and other informatics systems obeys the same physical laws as energy flow in steam engines and chemical reactions.

**V The Computing power of a physical device – The Landauer's Principle**

Thermodynamic considerations can be used to calculate the maximum speed of a processor from the power $P$ applied on it and its ambient temperature. In Turing model [12] erasing one bit and registering it again is an example of a logical operation. Therefore the bits rate $f$ of a file can be considered as the computing power of a physical device.

One can write the temperature of an emitter or a receiver as;

$$T = P/(k f \ln 2 ). \qquad (10)$$

Every physical system is surrounded by a thermal bath that emits thermal noise at a temperature $T_n$. The higher the bit rate, the lower the temperature of the file as the bit energy is reduced. Since the temperature of the file must be kept above the temperature of the noise $T_n$, namely $T > T_n$, the frequency has an upper limit. From Eq. (10) we conclude that $f < P/(kT\ln2)$ where $T$ should be about 10 times higher than the noise temperature. Therefore, the upper bound on computing power of any device is,

$$f \leq P/(10\, kT_n\ln2). \tag{11}$$

Namely the power applied on any computing device and its ambient temperature suffices to calculate a limit on its computing power. Von Neumann claimed that a computer operating at temperature $T$ must dissipate at least $kT\ln2$ energy per elementary act of information. In nature the ratio per nucleotide or amino acid is 20-100 $kT\ln2$ [13]. The minimum energy dissipation per logical operation as suggested by Von Neumann is known as Landauer's principle [14]. It is seen that it obtains naturally from the second law of thermodynamics.

**Summary**

When a random binary file is removed from an emitter or absorbed by a receiver, its energy may be considered as heat and its Shannon information as entropy. The average nat energy of the file is $kT$, where $T$ is shown to be the informatics temperature of the emitter or the receiver. If the binary file is not random, Shannon information is Boltzmann $-H$ function. This approach is shown to comply with the second law of thermodynamics, reproduces the Carnot efficiency and the Landauer's principle.

**Acknowledgements**: I thank Y.B.Band, R.D. Levine, J. Agassi and Y. Kafri for many useful discussions.